# Giant gate-tunable renormalization of spin-correlated flat-band states and bandgap in a 2D magnetic insulator

Pin Lyu[1,2]*, Joachim Sødequist[3]*, Xiaoyu Sheng[1]*, Zhizhan Qiu[1,2], Anton Tadich[4,5], Qile Li[4,6], Mark T. Edmonds[4,6], Jesús Redondo[7,8], Martin Švec[8], Thomas Olsen[3]†, Jiong Lu[1,2]†

[1]Department of Chemistry, National University of Singapore, 3 Science Drive 3, Singapore 117543, Singapore.

[2]Institute for Functional Intelligent Materials, National University of Singapore, 117544, Singapore.

[3]Department of Physics, Computational Atomic-scale Materials Design (CAMD), Technical University of Denmark, DK-2800 Kgs. Lyngby, Denmark.

[4]ARC Centre for Future Low Energy Electronics Technologies, Monash University, Clayton, VIC 3800, Australia.

[5]Australian Synchrotron, Clayton, Victoria, Australia.

[6]School of Physics and Astronomy, Monash University, Clayton, VIC 3800, Australia.

[7]Institute of Physics of the Czech Academy of Sciences, Cukrovarnicka 10, 162 00 Prague 6, Czech Republic.

[8]Faculty of Mathematics and Physics, Charles University, V Holešovičkách 2, 180 00 Prague, Czech Republic.

*These authors contributed equally to this work.

†Corresponding author. E-mail: tolsen@fysik.dtu.dk, chmluj@nus.edu.sg

**Abstract:**

Emergent quantum phenomena in two-dimensional van der Waal (vdW) magnets are largely governed by the interplay between the exchange and Coulomb interactions. The ability to tune the Coulomb interaction in such strongly correlated materials enables the precise control of spin-correlated flat-band states, bandgap ($E_g$) and unconventional magnetism, all of which are crucial for next-generation spintronics and magnonics applications. Here, we demonstrate a giant gate-tunable renormalization of spin-correlated flat-band states and bandgap in magnetic chromium tribromide ($CrBr_3$) monolayers grown on graphene. Our gate-dependent scanning tunneling spectroscopy (STS) studies reveal that the inter-flat-band spacing and bandgap of $CrBr_3$ can be continuously tuned by 120 meV and 240 meV respectively *via* electrostatic injection of carriers into the hybrid $CrBr_3$/graphene system, equivalent to the modulation of the Cr on-site Coulomb repulsion energy by 500 meV. This can be attributed to the self-screening of $CrBr_3$ arising from the gate-induced carriers injected into $CrBr_3$, which dominates over the opposite trend from the remote screening of the graphene substrate. Precise tuning of the spin-correlated flat-band states and bandgap in 2D magnets *via* electrostatic modulation of Coulomb interactions not only provides new strategies for optimizing the spin transport channels but also may exert a crucial influence on the exchange energy and spin-wave gap, which could raise the critical temperature for magnetic order.

**Introduction:**

Two-dimensional (2D) van der Waal (vdW) magnets with strong electronic correlations host a variety of exotic quantum phases, including skyrmion[1], quantum spin liquid[2] and quantum anomalous hall conductance[3], and thus these materials have attracted a vast amount of interest following the seminal reports on the layer-dependent magnetism in $CrI_3$[4] and $CrGeTe_3$[5] crystals. In the 2D limit, the Mermin-

Wagner theorem prohibits a spontaneously broken symmetry phase and thus long range magnetic order arises due to spin-orbit coupling[6], which explicitly breaks the spin rotational symmetry and leads to magnetic anisotropy. In addition, the interplay between the exchange interaction and strong electronic correlations determines the energy spectrum of the spin-correlated flat-band states as well as the bandgap, which play an essential role in spintronics applications. In particular, the presence of a desired bandgap combined with layer-dependent ferromagnetism enables the integration of 2D magnetic insulators into vdW heterostructures where they may act as tunnel barriers, replacing the traditional dielectrics for next-generation spintronics. Future progress in this field depends on the ability to tune the electronic correlations *via* electrical fields and control the spin-correlated band structures associated with the spin-selective transport barriers in magnetic tunnelling devices[7, 8].

Conventional strategies towards this goal include chemical[9] or electrochemical doping[10], which are often inflexible and can result in unintentional material degradation. In contrast, electrostatic gating offers a reversible and continuous modulation of the electronic and magnetic properties of the target system, providing excellent compatibility and flexibility for practical electronics and spintronics applications. Despite the recent successes in the experimental realization of electric control of magnetic ordering in 2D magnets[11, 12, 13], electrical modulation of spin-correlated flat-band states and bandgap values has not been demonstrated. Moreover, atomic-scale insights into the influence of gate-controlled carrier densities on electronic correlations remain elusive.

$CrBr_3$ represents an Ising-type ferromagnetic insulator with a Curie temperature of around 34 K[14] and has been proposed to comprise a correlated Mott–Hubbard-like insulator[15, 16, 17]. Here, we have taken monolayer $CrBr_3$ as one representative 2D magnetic insulator to probe the gate-tunable renormalization of spin-correlated flat-band states and bandgap using scanning tunneling microscopy (STM). As a direct study of insulating materials *via* STM is infeasible due to lack of a conductive path, one can circumvent this by integrating atomically thin 2D magnetic insulators with conductive layered materials, either by vdW technology[18, 19] or molecular beam epitaxy (MBE) growth[20, 21].

To this end, we present a STM/STS study of a magnetic $CrBr_3$ monolayer prepared by MBE on a gate-tunable graphene device. A direct comparison of STS and Angle Resolved Photoelectron Spectroscopy (APRES) measurements with spin-polarised Density Functional Theory (DFT) calculations enables us to assign each correlated state of $CrBr_3$ to individual spin-split manifolds, and to determine the bandgap of monolayer $CrBr_3$ on the graphene substrate. Moreover, we demonstrate a continuous tuning of $E_g$ by 240 meV and the inter-flat-band spacing between correlated states of $CrBr_3$ by 120 meV through the electrostatic gating. This is equivalent to the modulation of on-site Coulomb repulsion energy by 500 meV, as corroborated by our DFT calculations with Hubbard correlation. By directly probing the doping level of the graphene both underneath and near the $CrBr_3$ domain, we find that gate-induced carriers are injected into both $CrBr_3$ (self-screening) and graphene (substrate screening), yielding opposite effect on the screening. The substrate screening in graphene becomes weaker upon electron doping (reduced hole density), whereas the self-screening in $CrBr_3$ is enhanced. Our findings demonstrate the crucial importance of the screened Coulomb interactions in correlated materials and provide a new strategy for precise control of spin-correlated states in 2D magnets.

**Results:**

**Structure characterization of $CrBr_3$ monolayer on a back-gated G/h-BN device**

We carried out MBE growth of monolayer $CrBr_3$ on a back-gated device (Fig. 1a), consisting of a chemical vapor deposition (CVD) grown graphene monolayer placed on a 15nm-thick hexagonal boron nitride (h-BN) flake resting on a Si wafer with the 285 nm-thick oxide layer (the substrate is denoted as G/h-BN). In contrast to the $SiO_2$ substrate, the atomic flatness of h-BN dramatically reduces the surface roughness and charge inhomogeneity in graphene, providing an ideal platform for the sample growth and subsequent gate-dependent STM measurements. A direct evaporation of anhydrous $CrBr_3$ bulk powder at 400°C for 30 mins allows for the growth of monolayer $CrBr_3$ islands with an apparent height of ~ 0.63 nm (inset of Fig.

1c) on the back-gated G/h-BN device, as captured in the large-scale STM image (Fig. 1c). The closed-up STM image acquired at a sample bias of $V_S = 1.2\ V$ reveals the periodic triangular pattern with a characteristic lattice constant of ~ 0.64 nm (Fig. 1d). The triangular pattern can be correlated to three protruding neighbouring Br atoms in the upper layer (marked by orange spheres in atomic model in Fig. 1b), consistent with the pattern reported in the previous STM studies of CrBr$_3$[20, 21, 22].

**Electronic properties of CrBr$_3$ monolayer on the G/h-BN device**

We subsequently performed the differential conductance (d$I$/d$V$) measurements in order to probe the electronic properties of CrBr$_3$ on this G/h-BN substrate (Fig. 2b). The d$I$/d$V$ spectrum of as-grown monolayer CrBr$_3$ reveals two prominent peaks located at $V_S = 1.01\ eV$ (labelled as C$_1$) and $V_S = 1.69\ eV$ (labelled as C$_2$), and a shallow rise below the Fermi level (E$_F$) (labelled as V$_1$). To better determine the band edges of CrBr$_3$, we conducted d$I$/d$V$ measurement at an increased set-point (0.5 V, 200 pA), whereby the intensity of the valence band edge can be dramatically enhanced, turning the shallow rise into a much sharper one (Supplementary Fig. 1). A direct comparison of the d$I$/d$V$ spectrum taken over monolayer CrBr$_3$ on graphene and on bare graphene enables us to assign a sharp rise at $V_S = 0.44\ V$ to the conduction band minimum (CBM) and the kink (onset of the sharp rise) at $V_S = -0.85\ V$ to the valence band maximum (VBM), yielding a value for E$_g$ of 1.29 eV for monolayer CrBr$_3$ grown on a G/h-BN substrate (Supplementary Fig. 2, refer to the details of band edge determination in Supplementary Note 1). In addition, we observe the E$_F$ is closer to the CBM than the VBM, indicating n-doping of CrBr$_3$, presumably due to charge transfer from the graphene substrate.

We also note that the d$I$/d$V$ spectra taken from monolayer CrBr$_3$ grown on graphite[20] and NbSe$_2$ [22] substrates are rather featureless in the top valence band region (Supplementary Fig. 4). To further assist the identification of the VBM, we performed ARPES measurements of monolayer CrBr$_3$ to probe the top valence bands in the momentum space. Since ARPES measurements require a relatively large area of single-crystal domain, we used monolayer graphene on SiC (0001) as a suitable substrate to grow monolayer CrBr₃ for the measurements. The APRES measurement taken at $hv = 90\ eV$ resolves the valence band dispersion of CrBr₃ along the Γ–M direction of graphene (Fig. 2e). The Γ–M direction of graphene was chosen to eliminate the contribution from graphene's π-band (Dirac Cone) around the K-point, ensuring that the signal from CrBr₃ was clearly visible in the lower binding energy region (BE < 2 eV). We observe a faint band dispersion around BE= 1 eV (Fig. 2f), which is tentatively assigned to the highest occupied band near the VBM of CrBr₃. The onset of the VB is better visualised in the second derivative of the ARPES spectra and the VBM is determined to be 0.6 eV below the Fermi level through fitting of the energy distribution curve, which is slightly shallower than the value determined from the STS measurements. This difference can be attributed to the higher n-doping level of graphene on SiC ($E_D = -0.4\ eV$ for G/SiC[23] and $E_D = 0.12\ eV$ for G/h-BN), which facilitates charge transfer from graphene to CrBr₃. This presumably results in a larger band gap renormalization of CrBr₃, as will be discussed in detail below.

To better understand the electronic properties of CrBr₃, we performed spin-polarized band structure calculations using the Perdew-Burke-Ernzerhof (PBE) exchange-correlation functional on the Cr atoms (see methods for details). Each Cr atom is bonded to six Br atoms, which gives rise to an octahedral crystal field that splits the $d$-orbital of Cr atoms into $t_{2g}$ and $e_g$ manifolds. The spin degeneracy of the these two manifolds is further lifted by magnetic exchange interactions[24, 25], where the energy spacing between the majority and minority spin channels is governed by the effective on-site Coulomb interaction (U)[26] as illustrated in Fig. 2a.

As shown in Fig. 2c, the calculated density of states (DOS) of free-standing monolayer CrBr₃ reproduces the key features observed in the experimental d$I$/d$V$ spectrum (Fig. 2b). Specifically, the C₁ peak is derived from the spin-majority states ($e_{g,\uparrow}$) with equal contributions from Cr $d$-orbitals and Br $p$-orbitals, whereas the C₂ peak consists of spin-minority $t_{2g}$ states ($t_{2g,\downarrow}$) that mainly originates from Cr $d$-orbitals. The calculations reveal that the lowest bands of the unoccupied states are composed of spin-majority states ($e_{g,\uparrow}$) and shows a rather flat dispersion with the CBM located at the K point (Fig. 2d and Supplementary Fig. 5). In contrast, the VB shows a slightly more dispersive behaviour than the CB and the VBM is predicted to be located at the M point. Due to the large in-plane momentum, the electronic states at the M point exhibit a faster out-of-plane decay, which may explain the possible origin for a much-weaker d$I$/d$V$ signal of the VBM that can only be resolved at a reduced tip-sample distance in the STS measurement.

**Gate-tunable STS measurements of CrBr$_3$ monolayer on the G/h-BN device**

Our ability to gate the system allows us to study the fundamental issue of the importance of electron-electron interactions in a strongly correlated system. We do this by measuring how the d$I$/d$V$ spectra of CrBr$_3$ evolve as a function of the back-gate voltage (V$_g$), whereby carrier densities can be tuned continuously (Fig. 3b). Gate-dependent STS measurements show that the VBM undergoes a remarkable upward shift in energy from -0.89 ± 0.01 eV to -0.70 ± 0.01 eV, when V$_g$ is increased from -40 V to 40 V. In stark contrast, the CBM only shows a slightly downward shift from 0.46 ± 0.01 eV to 0.41 ± 0.01 eV (Supplementary Fig. 3). The opposite shift between CBM and VBM leads to a monotonic gap reduction from 1.35 ± 0.01 eV at $V_g = -40\,V$ to 1.11 ± 0.01 eV at $V_g = 40\,V$. Furthermore, the C$_1$ peak shows an upward shift from 0.96 ± 0.01 eV to 1.03 ± 0.01 eV, as V$_g$ is swept from -40 V to 40 V, while the C$_2$ peak moves downwards from 1.71 ± 0.01 eV to 1.66 ± 0.01 eV. This yields a gradual decrease in the inter-band spacing between C$_1$ and C$_2$ from 0.75 ± 0.01 eV to 0.63 ± 0.01 eV. Fig. 4a and 4c show the corresponding energy position of the VBM (cyan), CBM (magenta), C$_1$ (orange), C$_2$ (pink) peaks together with E$_g$ and inter-band spacing between C$_1$ and C$_2$ as a function of V$_g$. A significant bandgap renormalization of 240 meV is accompanied by a large modulation of the energy separation between the spin-correlated flat-band states (inter-band spacing between C$_1$ and C$_2$), which changes by 120 meV when V$_g$ is varied from -40 V to 40 V.

Next, we focus on deciphering the physical origin of the gate-tunable renormalization of the bandgap and spin-correlated flat-band states of monolayer CrBr$_3$. First, we are able to rule out the contribution from the Stark effect due to a negligible field-induced out-of-plane polarization in the 2D monolayer[27]. Second, electrostatic gating may result in charge redistribution, modulating the interfacial dipole, but such an effect can also be excluded because the change in the interfacial dipole only causes a rigid band structure shift and fails to explain the band renormalization observed here (Supplementary Fig. 6). We can also rule out the tip-induced bend bending effect (TIBB), as no systematic band edge shift observed here (Supplementary Fig. 7). We are thus left with the gate-tunable carrier density modulation as the key factor governing the renormalization. It has been previously demonstrated that giant bandgap renormalization in 2D materials can be realized by carrier doping, which modulates the Coulomb-hole self-energy and screened-exchange self-energy[28, 29]. Such effects have been experimentally observed in many 2D systems *via* alkali atom deposition[9, 30]. Apart from this, increasing the carrier density in the graphene substrate results in effective screening that increases the electron self-energy (quasiparticle stabilization)[31], which typically manifests itself as a quasiparticle bandgap reduction of the adjacent 2D semiconductors in graphene based vdW heterostructures[32, 33]. To determine which effect accounts for gate-tunable renormalization, we thus need to extract the gate-induced carrier density variation in both CrBr$_3$ and the graphene substrate.

One advantage of this study is that we can directly probe the gate-dependent carrier density of the graphene underneath (Fig. 3c) and close to the monolayer CrBr$_3$ islands (Fig. 3d) *via* d$I$/d$V$ spectroscopic measurements (Fig. 3a). At $V_g = 0\ V$, the Dirac point (E$_D$) of the graphene underneath CrBr$_3$ is determined to be 0.32 eV, suggesting a sizeable *p*-doping of graphene ($n_h = 1.12 \times 10^{13}\ cm^{-2}$). This is due to a significant charge transfer from graphene to CrBr$_3$ arising from the large work function difference between the two materials[34]. We noted that the E$_D$ of the graphene underneath CrBr$_3$ always lies above E$_F$, slightly decreasing from 0.34 eV at $V_g = -40\ V$ to 0.30 eV at $V_g = 40\ V$. This means that the hole carrier density in the graphene substrate undergoes a monotonic decrease from $1.25 \times 10^{13}\ cm^{-2}$ to $1.02 \times 10^{13}\ cm^{-2}$ within the accessible range of gate voltages (refer to the details for the determination of charge carrier in graphene in Supplementary Note 2). This effect therefore weakens the screening of the Coulomb interactions and is expected to enlarge the bandgap, opposite to the trend observed experimentally. In contrast, the $E_D$ of bare graphene close to the CrBr₃ domain varies from 0.23 eV at $V_g = -40\ V$ ($n_h = 5.85 \times 10^{12}\ cm^{-2}$) to -0.067 eV at $V_g = 40\ V$ ($n_e = 0.49 \times 10^{12}\ cm^{-2}$). The gate-induced carrier density variation ($\Delta n = 6.34 \times 10^{12}\ cm^{-2}$) on bare graphene is much larger than graphene underneath the CrBr₃ ($\Delta n = 2.32 \times 10^{12}\ cm^{-2}$), suggesting that more carriers are injected into the CrBr₃ layer (~$4.02 \times 10^{12}\ cm^{-2}$) (Supplementary Fig. 7). The gate-induced carriers injected into the CrBr₃ layer are expected to enhance the screening of the on-site Coulomb potential, resulting in a bandgap reduction. Therefore, the enhancement of screening resulting from an increase of electron density in CrBr₃ by $\Delta n = 4.02 \times 10^{12}\ cm^{-2}$ dominates over the reduced screening arising from a decrease of hole density by $2.32 \times 10^{12}\ cm^{-2}$ in graphene. The dominant screening effect from doped carriers in CrBr₃ can be rationalized by considering the effective length scale of carrier-induced screening from graphene and CrBr₃. Carriers injected into the flat conduction bands of CrBr₃ thus generates a more effective self-screening effect on the modulation of effective on-site Coulomb interactions than carriers injected into the graphene substrate.

Since CrBr₃ is a strongly correlated material wherein Coulomb interactions play a crucial role in determining its electronic properties, the renormalization of the bandgap and spin-correlated flat-band states caused by gate-induced carriers can be explained as a result of screened on-site Coulomb interactions. To model the local Coulomb interaction between the Cr $d$ electrons, we calculated the DOS of CrBr₃ on graphene with different Hubbard corrections (U). Fig. 4f shows the calculated DOS of CrBr₃ on graphene with different values of U separated into the two spin channels. We notice that the variation of U governs the spin splitting of the polarized manifolds, where smaller values of U lead to less spin-split manifolds[26]. In addition, a much larger splitting of the $t_{2g}$ manifold than the $e_g$ manifold is likely due to the fact that the $t_{2g}$ manifold has three sets of half-filled orbitals, whereas the $e_g$ manifold has only two sets of empty orbitals. The calculations reveal a monotonic shift of the $t_{2g,\uparrow}$, $t_{2g,\downarrow}$ and $e_{g,\downarrow}$ manifolds upon variation of U, while the $e_{g,\uparrow}$ manifold is pinned at $E_F$ due to the significant charge transfer from graphene to CrBr$_3$. As shown in Fig. 4a-d, a direct comparison of the experimental results with the calculated DOS reveals that the bandgap reduction with increased V$_g$ (increased carries density in CrBr$_3$) resembles the band structure evolution with the reduced U. By decreasing U from 0.5 eV to 0 eV, we observe a bandgap reduction of 140 meV, comparable to the experimental value of bandgap reduction achieved by ramping V$_g$ from – 40 V to + 40 V ($\Delta n = 4.02 \times 10^{12}\ cm^{-2}$). We thus conclude that the large bandgap renormalization observed here is mainly attributed to the gate-induced electron doping of CrBr$_3$, causing a decrease of U due to the enhanced screening of Coulomb interactions[35].

Apart from the bandgap renormalization, the modulation of U also results in a variation of the spin-correlated flat-band state spacing, namely the C$_1$-C$_2$ separation (the gap between $e_{g,\uparrow}$ and $t_{2g,\downarrow}$ derived flat bands). As illustrated in Fig. 2a, both the bandgap and C$_1$-C$_2$ separation correlate with the on-site Coulomb energy. Generally, the on-site Coulomb interaction governs the spin-splitting between majority and minority spin channels, whereas reducing the value of U results in a lower spin splitting. The spin splitting of the $t_{2g}$ bands is significantly larger than that of the $e_g$ band, situating the $t_{2g,\downarrow}$ manifold at higher energies than the $e_{g,\uparrow}$ manifold and this leads to a modulation of the bandgap defined by the distance between the $e_{g,\uparrow}$ and $t_{2g,\uparrow}$ manifolds. The large bare on-site Coulomb repulsion results in a much larger splitting of the $t_{2g}$ bands than the $e_g$ band, which also becomes the dominant effect in the determining the bandgap size and inter-band spacing.

As illustrated in Fig. 4e, for a large value of U (U = 0.5 eV, corresponding to $V_g < 0$ V), the $e_{g,\uparrow}$ and $t_{2g,\downarrow}$ manifolds exhibit a large separation. The $t_{2g,\downarrow}$ manifold contains multiple peak components arising from the splitting of the $d_{xy}$, $d_{yz}$ and $d_{xz}$ bands. With decreasing values of U, the $t_{2g,\downarrow}$ manifold moves downwards towards E$_F$ and its shoulder components start to mix with $e_{g,\uparrow}$ manifold. Hence, for vanishing U (U = 0 eV, corresponding to $V_g > 0\ V$), orbital mixing between the $e_{g,\uparrow}$ and $t_{2g,\downarrow}$ manifold gets enhanced and manifests itself as the emerging C$_1$ peak, leading to an upward shift of the C$_1$ peak with respect to the E$_F$. In addition, the C$_2$ peak shifts down and results in a significant reduction of inter-flat-band spacing by

340 meV. This is consistent with the trend of gate-tunable energy separation between $C_1$ and $C_2$ peaks observed experimentally.

**Discussion:**

In summary, we have demonstrated a gate-tunable bandgap and inter-flat-band spacing of $CrBr_3$ monolayer by 240 meV and 120 meV respectively *via* electrostatic gating. This giant renormalization of the band structure of monolayer $CrBr_3$ can be attributed to a dominant self-screening effect arising from the gate-induced carriers injected into $CrBr_3$ rather than graphene and amounts to the modulation of on-site Coulomb interactions by 500 meV. Our findings not only provide insight into the crucial impact of carrier densities in doped 2D magnets, but also offer new strategies for tuning the flat-band states in strongly correlated systems that are closely related to spin-polarised transport channels. Moreover, effective modulation of Coulomb interactions in 2D magnets may also significantly modify the exchange energy and spin-wave gap, which could result in higher critical temperatures for magnetic order.

**Materials and Methods:**

**MBE growth of $CrBr_3$ on Graphene/h-BN device:**

The Graphene/h-BN substrate was prepared by transferring monolayer CVD graphene onto exfoliated h-BN flakes that rest on a doped Si wafer coated with a 285 nm $SiO_2$ layer. The Graphene/h-BN substrate was then annealed at 400°C under $Ar/H_2$ gas in a tube furnace for 5 hours to remove possible polymer residues. Electrodes (Ti-3nm /Au-50nm) were deposited on graphene using the stencil mask technique. Prior to growth, the Graphene/h-BN device was degassed at 300 °C for 2 hours in UHV. Monolayer $CrBr_3$ islands were grown by the evaporation of anhydrous $CrBr_3$ bulk powder at 400°C, while the Graphene/h-BN device was held at 150°C for 30 mins.

**STM/STS measurements:**

The STM/STS measurements were conducted in an Omicron LT-STM held at 4.5 K

with a base pressure lower than $10^{-10}$ mbar. The STM tip was calibrated spectroscopically on Au(111) substrate. All the d$I$/d$V$ spectra were measured through a standard lock-in technique with a modulated voltage of 5 to 20 mV at a frequency of 812 Hz.

**ARPES measurements:**

The ARPES measurements were conducted at the Soft X-ray beamline at the Australian Synchrotron. Monolayer CrBr$_3$ was grown on Graphene/SiC substrate under a similar growth condition as on G/h-BN. All the measurements were performed at a photon energy of 90 eV from the beamline, and 21.2 eV using a Helium VUV source to ensure the best cross section, intensity and resolution. Full hemisphere Constant Initial State Angle Scans (CIS-AS) or so-called "azimuth" scans on Graphene/SiC at 21.2 eV and 90 eV were first performed at a binding energy of 1 eV below E$_F$ to determine the high symmetry directions of the Graphene/SiC surface. Energy distribution curves (EDC) were then taken along the corresponding high symmetry surface directions (Γ-M and Γ-K) from the Fermi level to 5 eV in binding energy to measure the valence bands of CrBr$_3$.

**DFT calculations:**

The DFT calculations have been carried out using the electronic structure package GPAW[36], which uses the projector augmented wave method and a planewave basis. The 2D CrBr$_3$ on graphene was simulated using a single unit cell of CrBr$_3$ with the experimentally determined lattice parameter and a $\sqrt{7} \times \sqrt{7}$ unit cell of graphene that were strained by 3.98% to match the lattice parameter. The structure was relaxed under the constraint of fixed unit cell until forces on all atoms were below 0.05 eV/Å. The electronic structure was calculated using the Perdew-Burke-Ernzerhof (PBE) exchange-correlation functional using a Γ-centered k-point grid with a density of 12 Å and a plane wave energy cutoff of 800 eV. Spin-orbit coupling was included non-self-consistently[37] in the calculated band structure shown in Fig 2d but does not have a significant influence on the electronic structure of CrBr3. The effective Hubbard correction used in this paper, $U_{eff} = U - J$, was applied to the $d$ orbitals of Chromium.

**Acknowledgements:**

J. L. acknowledges the support from MOE Tier 2 grants (MOE2019-T2-2-044 and MOE-T2EP10221-0005). T. O and J. S. acknowledge support from the Villum foundation Grant No. 00029378. Some of this research was undertaken on the Soft X-ray Spectroscopy beamline at the Australian Synchrotron, a part of ANSTO. We acknowledge the valuable discussion with Shawulienu Kezilebieke and Peter Liljeroth on the growth of CrBr3.

**Author contributions:**

J. L supervised projects. P. L fabricated the device, grew the sample and performed STM measurements and data analysis with the help of X. S; J. R. and M. S provided the graphene/SiC (0001) sample; P. L and X. S performed ARPES measurements with the help of A. T., Q. L, M. E; J. S performed the density functional theory calculation under the supervision of T. O; J. L and P. L prepared the manuscript with the contribution from J. S and T.O. All authors contributed to the discussion and helped in writing the manuscript.

**Competing interests:** The authors declare that they have no competing interests.

**Data and materials availability:** All data needed to evaluate the conclusions in the paper are present in the paper and/or the Supplementary Materials. Additional data related to this paper may be requested from the authors.

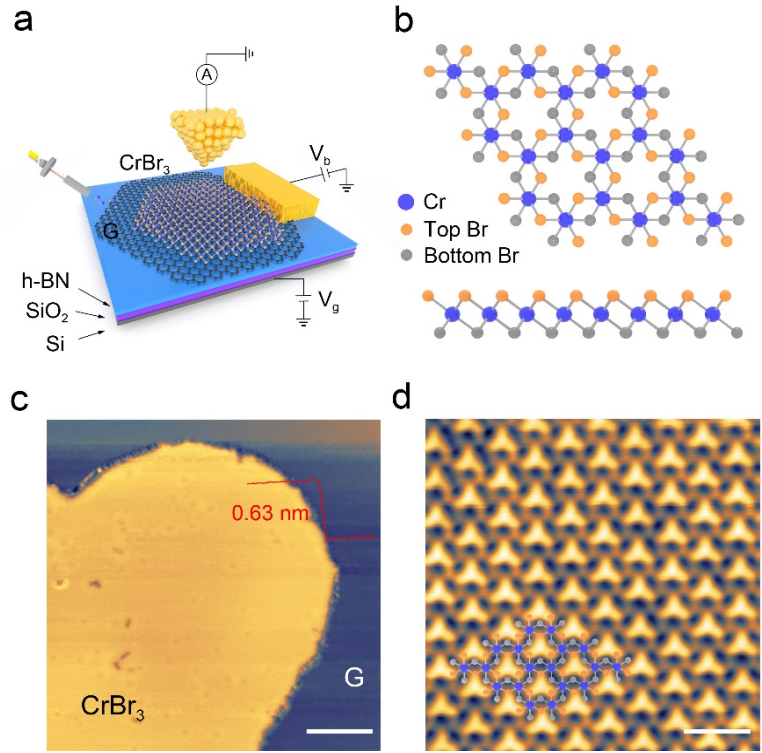

**Fig. 1. CrBr₃ monolayer grown on a back-gated G/h-BN device. a.** Schematic illustration of MBE growth of monolayer CrBr₃ islands on a back-gated graphene/h-BN device. **b.** Top and side view of the atomic structure of monolayer CrBr₃. The top layer of Br atoms is marked as yellow balls, forming a triangular shaped Br trimer. **c.** A representative large-scaled STM image of a monolayer CrBr₃ island. Inset shows the line profile along the domain edge. Scale bar: 10 nm. Setpoint: 1.5 V, 10 pA. **d.** Atomically resolved STM image of monolayer CrBr₃ with the overlaid atomic structure. Scale bar: 1 nm. Setpoint: 1 V, 10 pA.

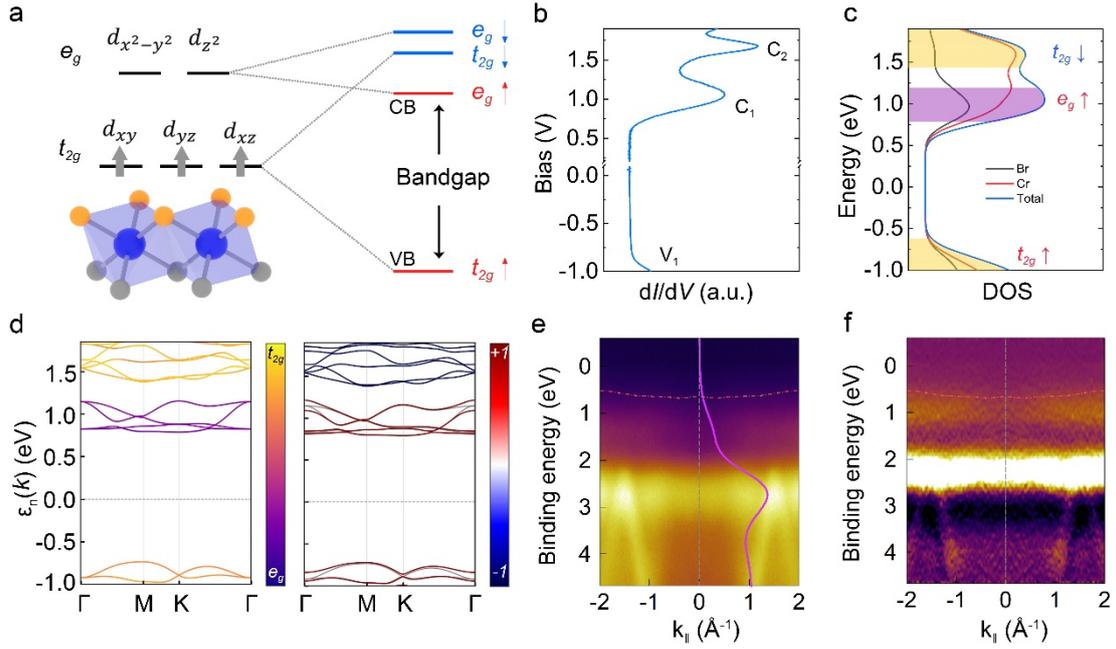

**Fig. 2. Electronic properties of monolayer CrBr3 on graphene. a.** Schematic illustration of the energy splitting of Cr $d$-orbitals into spin-polarised $e_g$ and $t_{2g}$ bands in the presence of octahedral crystal field and magnetic exchange interactions. **b.** Merged d$I$/d$V$ spectrum of monolayer CrBr3 at $V_g = 0\ V$. **c.** Calculated partial DOS of Br $p$-orbitals, Cr $d$-orbitals and total DOS. Note that the calculated DOS of CrBr3 is manually offset by 0.5 eV to align with the energy position of d$I$/d$V$ spectrum. **d.** Calculated band structure of monolayer CrBr3 using Perdew-Burke-Ernzerhof (PBE) exchange-correlation functional. The left panel shows the orbital information, where the yellow (purple) color denotes the $t_{2g}$ bands ($e_g$ bands) contributions; The right panel contains the spin information. The red (blue) color denotes the majority-spin (minority-spin) polarization along the out-of-plane direction. Note that SOC was included non-self-consistently in the right panel. **e.** Overall band structure along the Γ-M direction of graphene measured with ARPES at $hv = 90\ eV$. Inset: ARPES energy distribution curve. **f.** Second derivative of **e**. Red dashed line respresents the onset of the valence band. Note that the cone-like feature comes from the graphene/SiC saddle point located at the M-point.

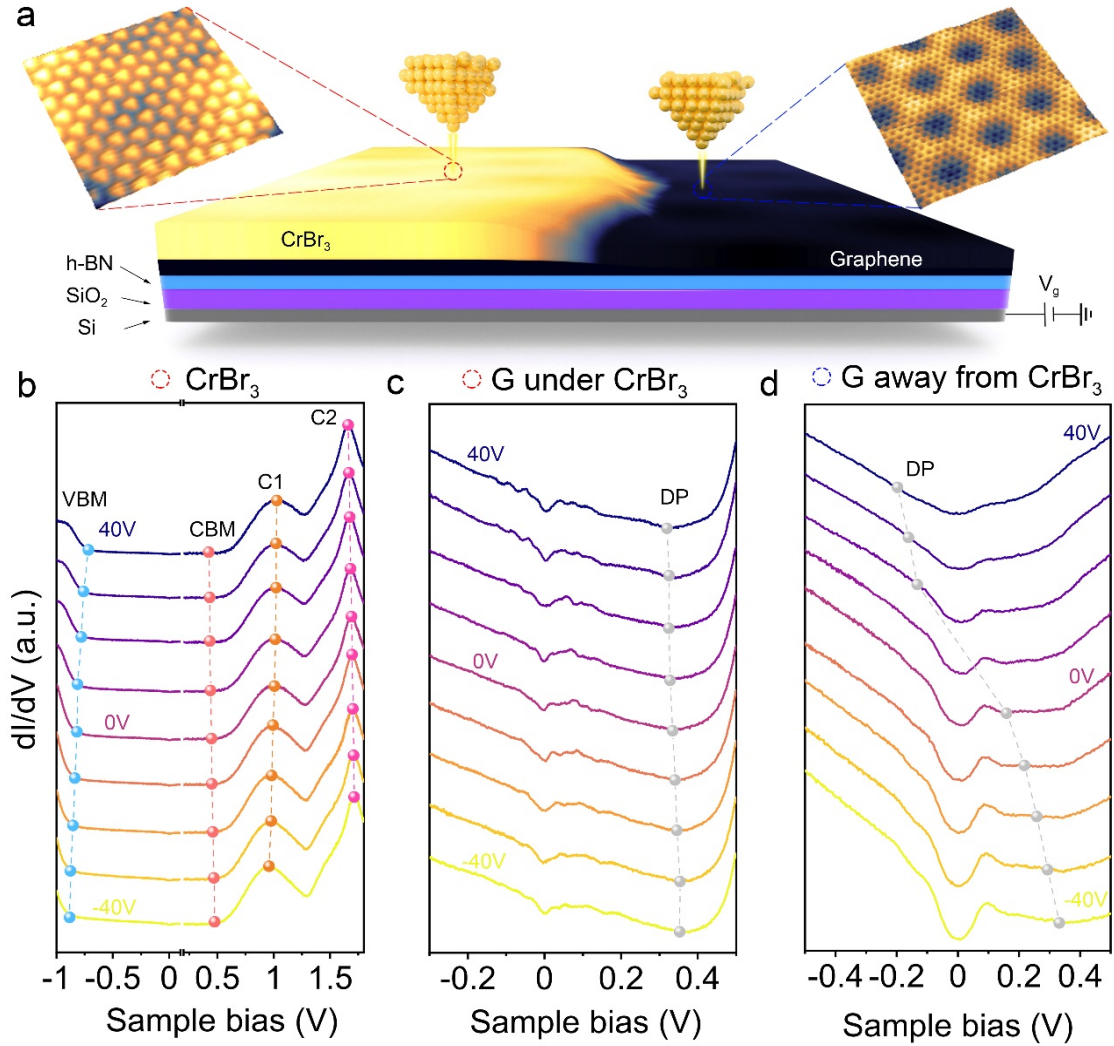

**Fig. 3. Gate-dependent d$I$/d$V$ spectra of CrBr$_3$ monolayer on graphene. a.** Schematic illustration of a region containing CrBr$_3$ domain and bare graphene. 3D STM images of CrBr$_3$ domian and bare graphene domain are included. **b.** Gate-dependent d$I$/d$V$ spectra of the monolayer CrBr$_3$ on graphene. The energy position of VBM, CBM, C$_1$ and C$_2$ peaks are indicated by cyan, magenta, orange and pink points, respectively. **c.** Zoom-in d$I$/d$V$ spectra taken at a closer tip-sample distance over the monolayer CrBr$_3$ on graphene to monintor the Dirac point position (marked as gray points) as a function of $V_g$. **d.** Gate-dependent d$I$/d$V$ spectra of a bare graphene surface nearby the monolayer CrBr$_3$ island (10 nm away from the edge of CrBr$_3$ island). The Dirac point is marked as gray points. At $V_g = 0\ V$, graphene/h-BN substrate shows a slightly $p$-type doping as the E$_D$ of graphene is located at 0.12 eV above the E$_F$.

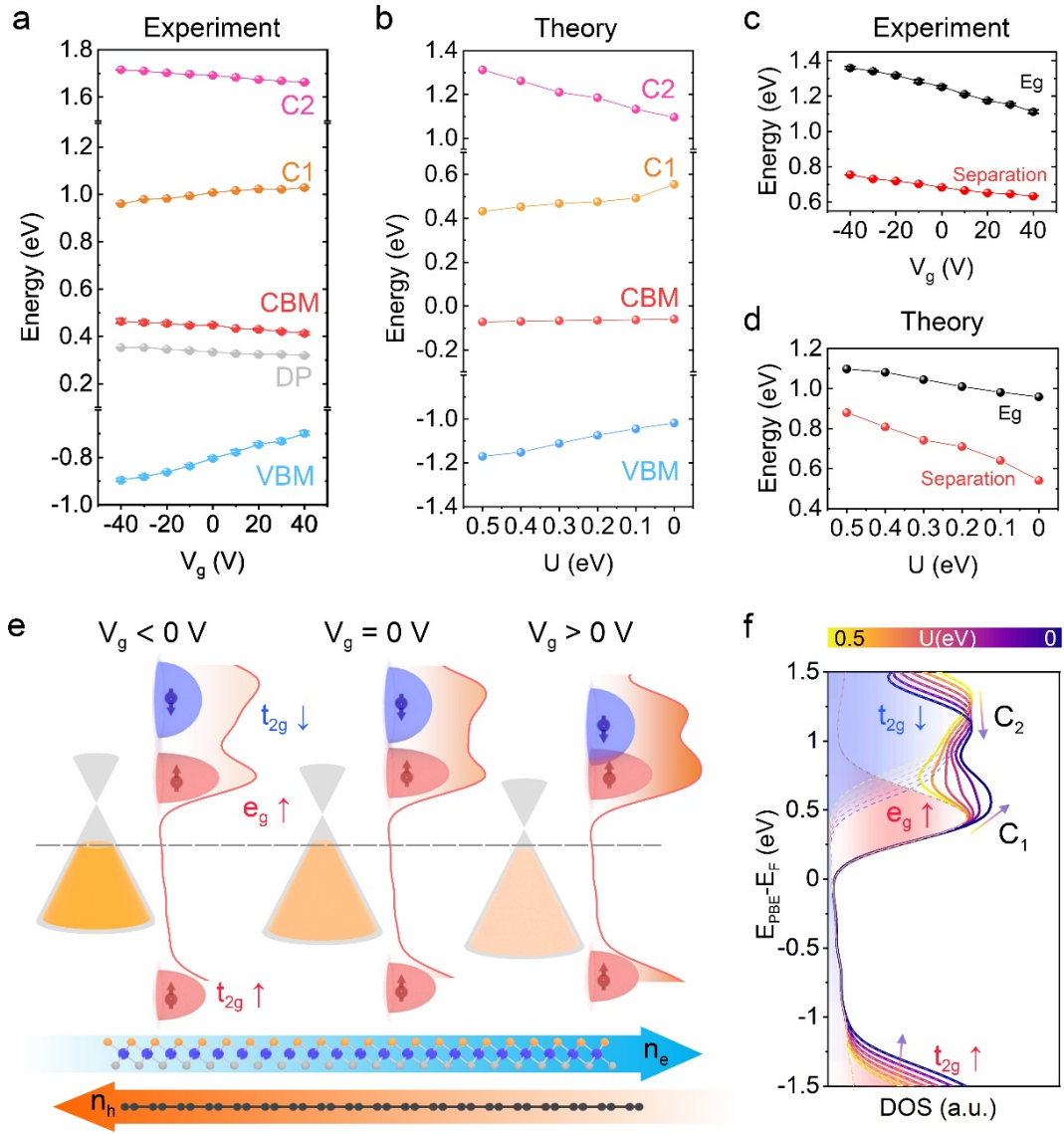

**Fig. 4. Gate-tunable bandgap renormalization and correlated flat-band states of monolayer CrBr₃ on graphene. a.** Energy position of the VBM (cyan), CBM (magenta), Dirac point (gray), $C_1$ (orange) and $C_2$ (pink) peaks as a function of gate voltage. **b.** Calculated energy position of the VBM (cyan), CBM (magenta), $C_1$ (orange) and $C_2$ (pink) peaks as a function of the on-site Coulomb repulsion (U). **c.** Measured bandgap and inter-flat-band spacing between the $C_1$ and $C_2$ peaks as a function of gate voltage. **d.** Calculated bandgap and inter-flat-band spacing between the $C_1$ and $C_2$ peaks as a function of U. **e.** Illustration of the band alignment upon the variation of $V_g$ and the corresponding U. **f.** Calculated DOS of monolayer CrBr₃ on graphene using different values of U. The $C_1$ and $C_2$ peak positions were determined

from the numerical maximum of the peak in the projected density of states and the VBM and CBM was obtained from the slope of the band edges in the projected density of states.